\documentclass[11pt,preprint]{aastex}
\usepackage{natbib,psfig}

\newcommand{\lumin}{{erg~s$^{-1}$}}

\newcommand{\msun}{\hbox{${M}_{\odot}$}}
\newcommand{\lsun}{\hbox{${L}_{\odot}$}}
\newcommand{\nh}{\hbox{${N}_{\rm H}$}} 
\newcommand{\simgt}{\lower 2pt \hbox{$\, \buildrel {\scriptstyle >}\over {\scriptstyle\sim}\,$}}
\newcommand{\simlt}{\lower 2pt \hbox{$\, \buildrel {\scriptstyle <}\over {\scriptstyle\sim}\,$}}

\newcommand{\cmsq}{\hbox{cm$^{-2}$}}

\begin{document}


\title{\boldmath$Chandra$-SDSS Normal and Star-Forming Galaxies I: 
X-ray Source Properties of Galaxies Detected by \boldmath$Chandra$ in SDSS DR2 }


\author{A.E.~Hornschemeier,$^{1,2}$ ~
T.~M. Heckman$^{1}$,A.~F. Ptak$^{1}$, C. A. Tremonti$^{3}$,
E.~J.~M. Colbert$^{1}$}

\altaffiltext{1}{Department of Physics and Astronomy, The Johns Hopkins University, 
3400 N. Charles Street, Baltimore, MD 21218, USA}

\altaffiltext{2}{Chandra Fellow}

\altaffiltext{3}{Steward Observatory, 933 N. Cherry Ave., Tucson, AZ 85721, USA}


\begin{abstract}

We have cross-correlated X-ray catalogs derived from archival $Chandra$ ACIS observations 
with a Sloan Digital Sky Survey (SDSS) Data Release 2 (DR2) galaxy catalog to form 
a sample of 42 serendipitously X-ray detected galaxies over the redshift interval 
$0.03 < z < 0.25$.  This pilot study will help fill in the ``redshift gap" between 
local X-ray-studied samples of normal galaxies and those in the deepest X-ray surveys.  
Our chief purpose is to compare optical spectroscopic diagnostics of activity 
(both star-formation and accretion) with X-ray properties of galaxies.  Our work 
supports a normalization value of the X-ray-star-formation-rate (X-ray-SFR) correlation
consistent with the lower values published in the literature.  The difference is in the allocation
of X-ray emission to high-mass X-ray binaries relative to other components such as hot gas, low-mass X-ray binaries,
and/or AGN.  We are able to quantify a few pitfalls in the
use of lower-resolution, lower signal-to-noise optical spectroscopy to identify X-ray sources (as 
has necessarily been employed for many X-ray surveys).  Notably, we find a few AGN that 
likely would have been misidentified as non-AGN sources in higher-redshift studies.  
However, we do not find any X-ray hard, highly X-ray-luminous galaxies lacking
optical spectroscopic diagnostics of AGN activity.  Such sources are members of the 
``X-ray Bright, Optically Normal Galaxy" (XBONG) class of AGN.

\end{abstract}





\section{Introduction \label{intro}}

X-ray emission in galaxies has at a basic level two origins:  gravitational
potential energy release from accretion processes and thermal emission from
hot gas.  Accretion processes may be divided among three main systems:  where a
lower mass ($\simlt1$~\msun), and correspondingly longer-lived ($>1$~Gyr), star accretes onto 
either a neutron star or black hole (low-mass X-ray binary; hereafter LMXB), where a
higher mass ($\simgt5$~\msun) and thus shorter-lived ($<10^{7}$~yr)  star accretes onto a neutron
star or black hole (high-mass X-ray binary; hereafter HMXB), and finally where
accretion occurs onto a nuclear supermassive (10$^6$--10$^{8}$~\msun) black hole 
(active galactic nucleus; hereafter AGN).   Hot gas is produced through a combination
of stellar mass loss and supernovae, and this phase is often the dominant one in the interstellar
media (ISM) of galaxies.   Indeed, one of the more notable results in X-ray studies of 
galaxies was finding the expected interstellar-medium (ISM) component of elliptical galaxies
\citep{Forman79}. The ISM in ellipticals is largely invisible at radio and optical wavelengths and
had for a time presented a mass-conservation problem given the expected output from stellar winds.
 The hot ISM in galaxies with younger stellar populations may be highly extended, as
supernova and massive stellar winds give rise to large-scale
outflows \citep[superwinds;][]{Heckman90}.
For the moment ignoring AGN, it is thus expected that in early-type galaxies, with 
older stellar populations, we will find that X-ray emission is dominated by LMXBs and
hot ISM.  In late-type galaxies, we expect a mix of emission from hot gas, LMXBs, and HMXBs
with starburst galaxies having a great amount of both hot gas and HMXBs.   The hot gas component
is typically X-ray soft (dominating at $\simlt1$~keV) whereas the X-ray binary emission is more 
X-ray hard (dominating the total X-ray emission at $>2$~keV).

It appears that, as expected, the properties of X-ray binaries are correlated with star 
formation rate (hereafter SFR) in many nearby ($\simlt 50$~Mpc) galaxies 
studied intensively with Chandra \citep[e.g., the X-ray Luminosity Functions of the binaries appear to
obey scaling laws based upon SFR; ][]{Kilgard02,ColbertXLF2003}.  
The total X-ray emission from binary systems may be written as a linear combination of the young population
associated with SFR and the older population associated with stellar mass  
\citep[i.e.,$L_{\rm X}=\alpha*SFR + \beta*M$; ][]{David92,ColbertXLF2003}.
In galaxies with a higher ratio of SFR to stellar mass, where the LMXB component is expected to
be negligible, the total X-ray luminosity may be dominated
by HMXBs and it may be possible to use the total X-ray luminosity, neglecting other contributions,
 as an SFR indicator
 \citep[e.g.,][]{Gilfanov2004SFR}.   Indeed, there have been indications that the total X-ray luminosity of a galaxy
may serve as a star-formation diagnostic at higher redshift ($z\simgt0.3$,
$d\simgt1700$~Mpc) \citep[e.g.,][]{GrimmSFR,BauerXII}.  Of cosmological interest, it also appears
that X-ray emission from distant Lyman Break galaxies may be similar to galaxies in the nearby
Universe \citep[e.g.,][]{BrandtLyBreak,Seibert2002,Nandra02,Reddy2004}. 

The X-ray band holds much promise as an independent probe of SFR.  For instance, it has been
shown that one of the chief uncertainties in the determination of the power-law index 
of the Initial Mass Function (hereafter IMF) arises mostly from unresolved binary star 
systems \citep{Kroupa2001}.  Theoretical models of the production of accreting X-ray binary systems,
which are probed through X-ray observations of star-forming galaxies, are 
sensitively dependent upon properties of the binaries in galaxies \citep[e.g.][]{Sipior2004}.  It is
thus possible that as both the models and the empirical X-ray/SFR relations improve we may develop
a probe that is particularly sensitive to the binary populations in galaxies. Interesting
conclusions regarding the evolution of the ``binary IMF" have been discussed in \cite{NormanXLF}.

Constraining the X-ray/SFR relation requires understanding the X-ray components of galaxies 
that are not directly associated with current star formation.  There have been fairly different 
approaches to this problem.   For instance, \cite{GrimmSFR} select galaxies 
with high ratios of SFR to stellar mass (hereafter SFR/M$_{*}$) claiming that they can then reject the LMXB component more easily.  \cite{GrimmSFR} claim that the ${L_{\rm X}}$-${\rm SFR}$  
relation is linear only at higher SFR ($\simgt 5$~\msun~yr$^{-1}$) 
where high-mass X-ray binaries (HMXBs) dominate the X-ray emission in galaxies.
At lower SFR, they claim that the HMXB  ${L_{\rm X}}$-${\rm SFR}$
relation becomes non-linear due to small number statistics \citep{Gilfanov2004stats}.  
Four other groups have found that the ${L_{\rm X}}$-${\rm SFR}$ relation is linear from low to
high SFR \citep{BauerXII,Ranalli03,ColbertXLF2003,Persic2004}.  
Both \cite{Persic2004} and \cite{ColbertXLF2003} include galaxies with 
lower SFR/M$_{*}$ values and make direct accommodation for the LMXB component.   
Additionally, \cite{ColbertXLF2003} uses only the resolved point source population
so that the hot gas and nuclear (AGN) component may be rejected. \cite{Persic2004} use X-ray spectral 
fits to reject emission from hot gas and AGN.   \cite{BauerXII}, whose work is at higher redshift,
select optical emission-line galaxies which do not show signs for AGN activity.  
The result is that there is also quite a spread 
in the {\it normalization} of the ${L_{\rm X}}$-${\rm SFR}$ relation; 
published values differ by approximately a factor of five.  This difference appears to lie with the
way in which one treats the LMXB and AGN X-ray ``contamination" in the total X-ray luminosity of galaxies.

Understanding this relationship is also quite important for the study of distant galaxies, 
where AGN contamination may be much more difficult to constrain due to the dilution of 
AGN optical spectral features by the host galaxy light \citep[e.g.,][]{Moran2002}.
High-$z$ classifications are also more difficult due to the inaccessibility 
of the H$\alpha$ emission line, a useful discriminator between true 
absorption-line dominated (passive) galaxies and those which contain highly obscured AGN
\citep[e.g.,][]{Goto2003}.  There is even an extreme class of objects
referred to as ``X-ray Bright, Optically-Normal Galaxies" (XBONGS) which have 
optical absorption-line dominated spectra but very high hard X-ray luminosities \citep{Comastri02}.
Any census of AGN activity in the optical waveband must be concerned with the frequency of XBONG-type
objects.

Systematic methods for separating X-ray-detected galaxies dominated by AGN emission from 
those dominated by starburst emission have recently seen a breakthrough with e.g.,
the Bayesian analysis of \cite{NormanXLF}.  This work has allowed the first normal/starburst galaxy
X-ray luminosity function to be constructed at $z\approx0.3$ and $z\approx0.7$ \citep{NormanXLF}.  
There is clear (and expected) evolution in this XLF, which has a lognormal shape,
similar to FIR luminosity functions.  The evolution appears to be consistent
with the expected $(1+z)^{2.7}$ evolution of SFR.  As such high-$z$ normal galaxy XLF studies 
continue to improve, we may hope to detect subtle evolutionary differences between e.g., the X-ray and FIR in
order to constrain e.g., X-ray binary evolution over long timescales.  Such work critically depends upon a
thorough understanding of the relative contributions of X-ray emission associated with SFR, AGN or quiescent
populations in galaxies.

There remains a problem:  there still is no suitable normal galaxy sample
constructed in the nearby ($z\simlt0.1$) Universe with which to make comparisons with 
high-redshift X-ray work \citep[e.g.,][]{HornOBXF,NormanXLF}.  
The observational expense of studying extremely local ($d<100$~Mpc) galaxies to sufficient depth
to probe both diffuse X-ray gas and the binary LF (down to $\sim10^{38}$~\lumin, thus reaching 
neutron-star binaries) is great;  exposures of $~50$~ks are typically required. 
Currently there are no sufficiently large ($>30$ galaxies) samples 
complete in any particular observational property (optical magnitude and/or
morphology, starburst properties like the ratio of SFR/mass, etc.).
Also, while the range of SFR sampled in X-ray-SFR studies to date has been quite large,
it has in some cases been heavily biased towards higher star-formation rates (SFR).  The studies
aforementioned have typical average SFRs of
$\approx10$~M$_\odot$~yr$^{-1}$, going up to $\approx100$~M$_\odot$~yr$^{-1}$.  This makes
probing the possible transition to non-linear ${L_{\rm X}}$-${\rm SFR}$ quite difficult.
Also, there have been heterogeneous methods used to measure
stellar mass and SFR.  For instance, \cite{GrimmSFR} were forced to use approximately 10 separate FIR
studies of nearby galaxies in order to derive SFRs, and for the
same galaxies the FIR flux estimate often differed by factors of several.   Reliable
UV, H$\alpha$, and radio SFR estimates are not always available for the
same galaxy.  This is expected to change dramatically with 
well-observed nearby-Universe samples such as the Spitzer Infrared Nearby Galaxies Survey
\citep{KennicuttSINGS}.
Finally, there is some concern about ``distance-distance"
biases introduced by the large range in distances sampled by some of these studies.  For instance,
many studies place high-$z$ galaxies ($z>0.8$) from the $Chandra$ Deep Fields on the same correlation
fits as galaxies in the nearby Universe ($d<50$~Mpc).  However, strong bias is only expected when the range
in flux is quite small whereas the distance range is large;  for most studies this is not a serious problem.

 It is thus critical to observe galaxies with moderate SFR values 
($\sim1$--10~M$_\odot$~yr$^{-1}$) as well as galaxies with no evidence for 
current star formation.   Since the observational challenge at higher redshift is
separating the type-2 AGN from the normal/star-forming galaxies, 
low-to-moderate luminosity type 2 AGN should also be included.  The sample should have
{\it uniformly} measured SFRs and stellar masses and some highly accurate diagnostic of 
accretion activity.  With these requirements in mind, we have turned to the Sloan Digital Sky
Survey (SDSS) Data Release 2 (DR2) catalog and the $Chandra$ X-ray Observatory archive to search for a sample of X-ray-detected 
galaxies near $z\approx0.1$.   DR2 is the second major survey-quality data release of the SDSS
\citep[April 2004; ][]{Abaz2003}.

At $z\simgt0.05$ the space density of galaxies with $R\simlt18$ reaches $\approx100$~deg$^{-2}$, 
so that one would expect $\approx7$ galaxies in each $Chandra$ ACIS-I field of view 
($\approx16^{\prime}\times16^{\prime}$).  The SDSS is covering
$\approx10$,000 square degrees to this optical depth.  The typical SDSS galaxy redshifts 
($z\approx0.1$) are local enough to provide a calibration sample for the high-redshift 
studies of galaxies.  We are working towards providing the
missing calibration of the X-ray properties
of galaxies at $z\approx0.1$ by studying a uniform sample
of SDSS galaxies with serendipitous $Chandra$ detections.

The major advantage of the SDSS is its uniformity; all galaxies have been observed with
the same telescope and spectrograph; there is no ambiguity about optical filters or
other calibration issues.  The spectroscopic coverage of galaxies in the
SDSS DR2 over its $\approx2$,600 square
degree area is complete down to $R\approx17.8$ \citep{Stoughton02}.
$Chandra$ has the sensitivity to reach galaxies with $\log{({{f_{\rm X}}\over{f_{\rm R}}})} \approx -2$
(typical of e.g., luminous dusty starbursts, Alexander et al. 2002) in the
0.5--2~keV band at $r\approx18$ with 40~ks of ACIS exposure.\footnote{The $r$ magnitude
here is the SDSS $r$-band filter.}
Thus the depths of coverage of many $Chandra$ ACIS observations
should be sufficient to place significant constraints on the galaxies with SDSS spectra.
\nocite{davoXI}

Our work on individually X-ray detected SDSS galaxies is quite complementary to existing work.    
There have been some initial promising results on spiral galaxies from statistical 
analysis of galaxies (not individually detected) using a combination of shallow 
($\simlt 5$~ks) $XMM-Newton$ observations and the 2dF redshift survey \citep{Georgakakis03}.   
\cite{Georgakakis04needles} have also carried
out a focused program to identify optically bright, X-ray faint objects which are 
typically quiescent galaxies \citep[e.g.][]{HornOBXF}, reporting on constraints on 
X-ray galaxy number counts at bright X-ray fluxes using SDSS and $XMM$.   Our study only selects
against Type-I AGN and unresolved sources, we make no exclusion of sources with higher X-ray-to-optical
flux ratios (which may include some dusty starburst galaxies and higher-mass elliptical galaxies).
Our focus on {\it optically}  selected, individually X-ray detected galaxies also complements 
other serendipitous $Chandra$ programs which are driven by X-ray selection and naturally favor  
AGN \citep[e.g., the $Chandra$ Multiwavelength Project, ChaMP; ][]{Green03}.

The purpose of this paper is to compare optical activity diagnostics such as emission-line
ratios with the X-ray properties of galaxies.  The ultimate goal of our studies will be to assemble
a complete field galaxy sample; this paper is a pilot study which works towards that ultimate goal 
(in fact, although we avoid cluster centers, a significant fraction of the galaxies are in cluster environments).  Our chief emphasis is on the X-ray/SFR relation, but we also explore the X-ray properties
of quiescent galaxies and AGN.  A future paper will cover the normal galaxy SDSS XLF.

The cosmology assumed in this paper is $H_0=70$~km~s$^{-1}$ Mpc$^{-1}$, $\Omega_{\rm M}=0.30$,
$\Omega_{\Lambda}=0.70$.


\section{Sample Construction}

\subsection{X-ray data}
\label{xrdata}

Our primary goal is to characterize the X-ray emission from star-forming (non-AGN) 
galaxies in the SDSS DR2 database.  To accomplish this, we require $Chandra$ sensitivity to 
detect such galaxies at the median SDSS redshift of $z\approx0.1$.   We only consider 
$Chandra$ observations of at least 30~ks, which ensures that we reach 0.3--8~keV X-ray luminosities
of $\approx2.7\times10^{40}$~\lumin (assuming $\Gamma=1.8$).  Based on the \cite{Ranalli03} 
SFR/X-ray relation, which is for the {\it total} X-ray luminosity,
 this would correspond to a SFR sensitivity of at least 6 $\msun$~yr$^{-1}$. 
 $Chandra$ observations significantly shorter than this would not be useful for our purposes.

\begin{figure}[t!]
\centerline{\includegraphics[scale=0.80,angle=0]{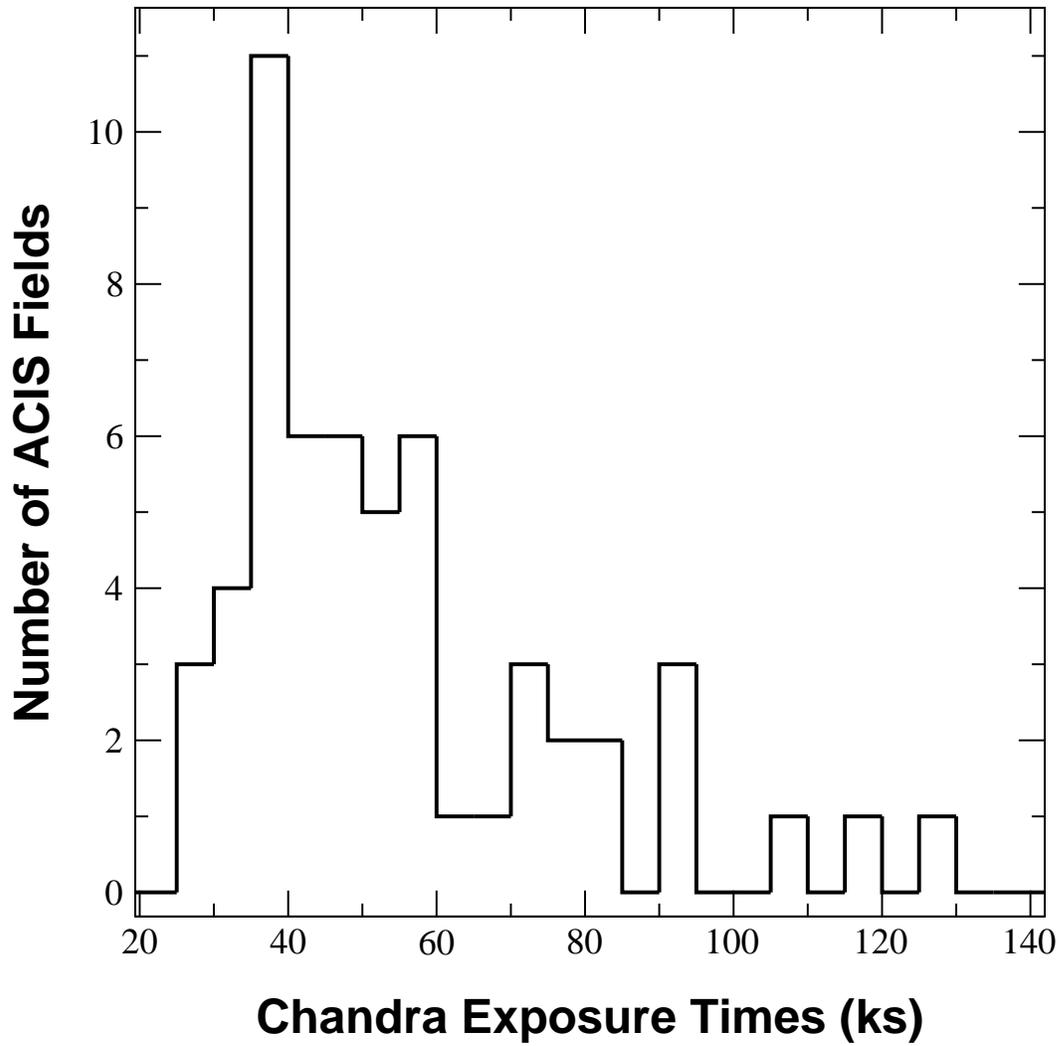}}
\caption[]{\label{XRexposures} X-ray exposure times for the 53 $Chandra$ ACIS fields in this study.  There are three fields at or
below the 30~ks lower cut-off because the net effective exposure time was slightly shorter than the nominal observation time (which
was used to query the archive).}

\end{figure}

We queried the $Chandra$ archive for ACIS non-gratings observations of at least 30~ks depth 
as of May 2004. We then used various SDSS cross-correlation and finding chart tools 
(publicly available through the main SDSS website\footnote{The SDSS SkyServer website is
http://cas.sdss.org/. } to ascertain that 53 fields fell within the SDSS DR2 spectroscopic
footprint.  Figure~\ref{XRexposures} gives the distribution of exposure times 
for these 53~fields;   24 of the fields had X-ray detected SDSS galaxies in them.  

These 53 X-ray datasets were reduced using the {\sc XASSIST}\footnote{See http://xassist.pha.jhu.edu}
 suite of software tools  \citep{Ptak2003}.  
{\sc XASSIST} uses the basic data reduction steps recommended by the CXC
``threads."  The data are reprocessed to level 2 event files
starting from the level 1 events and useful auxiliary products such as exposure maps
are created.  The optional 0.5 pixel position randomization was not performed. After  
reprocessing, each CCD is treated as a different detector and
individual CCD data are processed separately. For data processing, PI
values were constrained to be in the range 14--548, corresponding to the
energy range 0.3--8.0 keV.  

Sources are detected using the CIAO routine
{\sc WAVDETECT} \citep{Freeman02}, with the detection probability set at $1
\times 10^{-6}$.  Background flares were removed by fitting the background
light curve with a constant and excluding times where the background level
deviated by more than 10$\sigma$ from the mean (such a conservative limit
was chosen since the background is typically negligible for point-like
Chandra sources).  An image extracted for each source was then fit with an
elliptical Gaussian model plus sloping background model using Poissonian
statistics (i.e., the Cash ``C" statistic). This provides an improved
estimate of the extent and total count rate.  The significance of each
source is then re-assessed based on the total source counts and local
background derived from the spatial fit. Sources with Poisson
probabilities below the 95\% level are deleted from the source list.
Response matrices are then created for each source to estimate the source flux
assuming a power-law model with $\Gamma=1.8$ and the Galactic $N_H$ value
at the local of the source (determined using the FTOOL ``nh").  We then visually 
inspected the sources to look for problems such as proximity to CCD chip gaps (resulting
in some re-processing) and/or proximity to bright X-ray sources (a few sources were rejected).

As a sanity check, we compared the X-ray catalog produced by {\sc XASSIST} with the published X-ray
catalog for the $Chandra$ Deep Field-North \citep[CDF-N][]{davocatalog}, and found that of the 160 X-ray sources
identified by {\sc XASSIST}, 138 have matches in the deeper CDF-N catalog.  The ones that do not have matches in many 
cases show signs of X-ray variability with higher S/N in the individual shorter observations.   The astrometry in the two 
catalogs matched to within $\approx1$\farcs5.  This offset may easily be explained as the difference
between the astrometric corrections of \cite{davocatalog} and our ``raw" $Chandra$ coordinates.

\subsection{Optically-derived Galaxy Parameters from SDSS DR2}

The X-ray observed galaxies in this sample were drawn from 
 the SDSS DR2 `main' galaxy sample \citep{Strauss2002}.   
Details 
concerning the spectral characterization (emission-line measurement, etc.), stellar 
mass estimation, and star-formation rate estimates may be found in \cite{Kauff2003mass}, 
\cite{Kauff2003AGN}, \cite{Brinch2004} and \cite{Tremonti2004}.  These high-quality measurements were made 
for 211,894 galaxies and are now publicly available \citep[a synopsis of the data
products and of the project is given in ][hereafter ``the SDSS DR2 sample"]{BrinchDR22004}.
This group made made a number of improvements over the standard SDSS 
spectroscopic pipeline in e.g., emission-line flux measurements from continuum-subtracted spectra.

\begin{figure}[t!]
\centerline{\includegraphics[scale=0.80,angle=0]{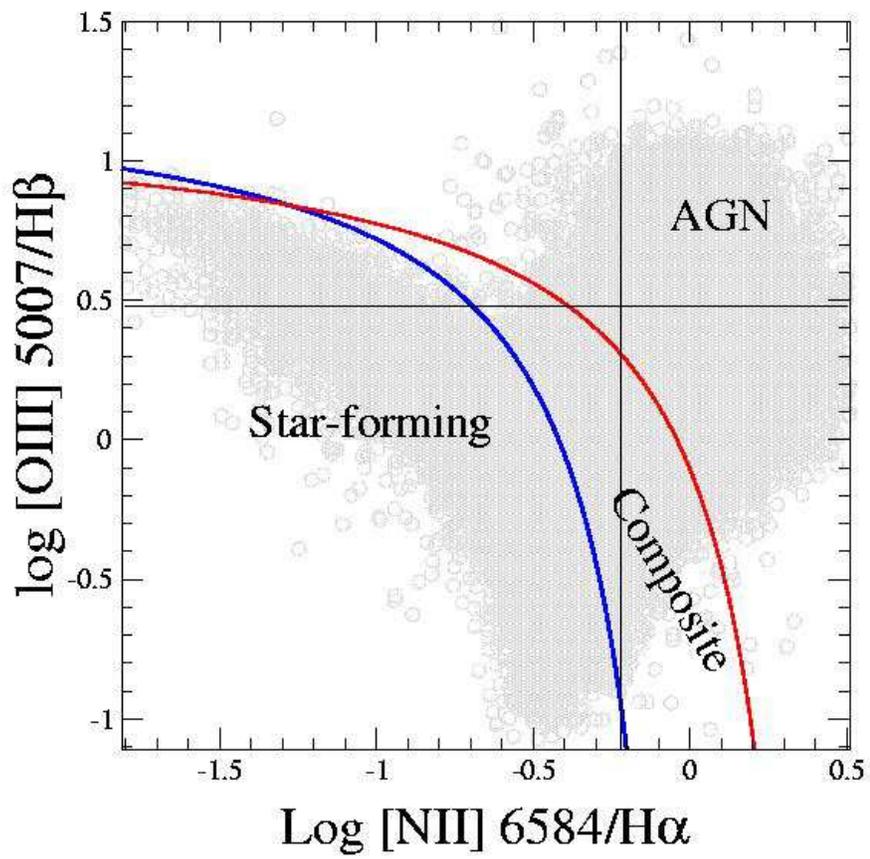}}
\caption[]{\label{BPTplain}  Emission-line ratios for the total SDSS DR2
sample of \cite{BrinchDR22004}.  }

\end{figure}

The standard emission-line ratios (e.g., $[NII]/H\alpha$ and $[OIII]/H\beta$)  were used to 
characterize the galaxy spectra; such line ratio diagrams are often referred to as ``BPT diagrams"
after \cite*{BPT}.  Higher values of $[NII]/H\alpha$ ($>0.6$) generally 
indicate AGN activity, with $[OIII]/H\beta$ discriminating between Low Ionization 
Nuclear Emission-line Regions \citep[LINER;][]{Heckman80} and Seyfert-type objects 
(LINERs have $[OIII]/H\beta<3$).  \cite{Kauff2003AGN} have suggested two lines of demarcation 
(see Figure~\ref{BPTplain}) between AGN and starburst galaxies in the BPT diagram 
which broadly separate galaxies into three classes:  AGN, star-forming 
galaxies, and AGN/star-forming composites.  The top line is the theoretical upper bound 
for starburst galaxy emission-line ratios from \cite{Kewley01}.  Galaxies with values to the right (above) 
this line in the BPT diagram are expected to have appreciable 
contribution from an AGN to their emission-line flux.  The second boundary, below which AGN 
activity is not expected and galaxies are typically star-forming galaxies, is a modified 
version of the \cite{Kewley01} line given by (Kauffmann et al.~2003a; see Figure~\ref{BPTplain}).
The optical classification of galaxies between these lines is AGN-starburst ``composite".  
Composite galaxies are expected to have $\approx41$\% of their [OIII] luminosity density and
$\approx 11$\% of their H$\alpha$ luminosity density arising from an AGN \citep{Brinch2004}.

One problem encountered in any star formation study is that the SFR
cannot be measured directly.  A large number of star formation indicators
have been theoretically or empirically calibrated at wavelengths ranging
from the UV to the radio.  In this work we adopt the total (i.e.
aperture-corrected) SFRs of \cite{Brinch2004}. These SFRs are
derived from the optical nebular lines (principally H$\alpha$) for galaxies
classified as star-forming, and from the 4000\AA ~break for galaxies
classified as AGN.  The majority of previous X-ray/SFR studies have relied
on far-infrared (FIR) SFR estimates.  However, \cite{Kewley02}
demonstrated that H$\alpha$ and FIR SFR estimates agree to 10\% when proper account is
made for attenuation.

For a full description of the SFRs the reader is referred to \cite{Brinch2004}; we briefly summarize the methodology here. The star formation rate within the
SDSS fiber apertures is estimated from the strong optical emission lines
([OII], [OIII], H$\beta$, H$\alpha$, [NII], [SII]) using a Bayesian approach and a grid
of $2 \times 10^5$ nebular models \citep{Charlot2001}.  The SFR is estimated simultaneously with the
metallicity, ionization parameter, dust attenuation, and dust-to-metal
ratio in order to account for variations in the unattenuated H$\alpha$/H$\beta$
ratio and the conversion between H$\alpha$ luminosity and
SFR.  This approach differs from the standard Kennicutt (1998) translation
between H$\alpha$ and SFR but there is good agreement in the mean conversion factor 
(see Figures 7 and 8 of Brinchmann et al. 2004b).  For galaxies harboring an AGN, the SFR
in the fiber is estimated from the D4000 spectral index, a measure of the
4000~\AA ~break.  This secondary SF indicator is calibrated consistently from
the star-forming galaxy sample.  Because the SDSS fibers only encompass
$\sim1$/3 of the galaxy light, the SFRs must be aperture corrected.  The galaxy
colors outside of the fiber aperture are used to estimate this correction.
Brinchmann et al. (2004b) adopt a Kroupa IMF \citep{Kroupa2001}.  To convert to
a \cite{Salpeter1955} IMF between 0.1 and 100 \msun, one would multiply the
Brinchmann SFRs by a factor of 1.5.  In this paper we correct the SFRs
of our comparison sample (see \S \ref{IMFcomp}) to the Kroupa IMF.

\subsection{SFR measurements of the comparison studies}
\label{IMFcomp}

There are five studies which we compare in this work (in order of discussion): 
\cite{ColbertXLF2003}, \cite{Persic2004}, \cite{BauerXII}, \cite{Ranalli03} and 
\cite{GrimmSFR}.  All of the SFR measurements may be traced back to the FIR-SFR 
calibration of \cite{Kennicutt98}, which is based upon the ``standard" 
Salpeter IMF discussed above.  However, in this section we aim to quantify the differences and
uncertainties in order to uncover any real disagreement among the various X-ray/SFR
relations.

\cite{ColbertXLF2003} and \cite{Persic2004} used \cite{Kennicutt98} FIR-SFR conversions. 
However, both made some additional correction to the IRAS FIR luminosity.
 \cite{ColbertXLF2003} added  the ultraviolet luminosity to the IRAS FIR luminosity to 
produce a ``corrected" FIR luminosity (which is higher than the directly measured FIR by up to a factor
of 2) and then apply a bolometric correction of a factor of
1.4 (thus their FIR luminosities may be up to 40\% higher than IRAS FIR luminosities).   
\cite{Persic2004} note that at lower SFR, non-SF-related emission such as cirrus may be present and
use the $B$-band luminosity to make this correction (a ``minor" 14\% correction).   
Therefore, both these studies (Colbert and Persic) are based
upon IRAS fluxes with a 14--40\% correction factor applied.

\cite{Ranalli03} used the \cite{Kennicutt98} FIR-SFR conversion (based upon the standard Salpeter
IMF) and the \cite{Condon1992} radio-SFR conversion.  Inspection of the Ranalli equations indicates consistency
with the \cite{Kennicutt98} calibration despite the  
\cite{Condon1992} radio-SFR conversion factor being for high-mass stars ($>5$~\msun) rather than for the more
inclusive mass range for the Salpeter IMF (0.1--100~\msun).  

\cite{BauerXII} use the radio-SFR relation of \cite{Yun01}, who derive the FIR-radio correlation
and then use the \cite{Kennicutt98} FIR-SFR relation to derive a radio-SFR conversion.
 The \cite{Yun01} SFR-radio normalization is lower than
the \cite{Condon1992} normalization in the sense that Yun predicts {\it less} radio emission per unit
SFR by a factor of $\approx2.3$.   Since the Yun SFRs are for the standard Salpeter IMF over a 0.1--100~\msun
interval but the \cite{Condon1992} SFRs are for a narrower mass interval ($>5$~\msun only), this difference in the
calibration of the 1.4~GHz estimator is as expected.
Thus, \cite{BauerXII} uses a 1.4~GHz estimator that appears consistent with the
rest of the FIR measures and with the standard Salpeter IMF.

\cite{GrimmSFR} is somewhat more difficult to compare with the others.  The FIR-SFR conversion used 
is the same Kennicutt (1998) conversion used by the others, but their measurements are much more
heterogeneous than the other authors.  \cite{GrimmSFR} have searched the literature for SFR 
measurements in nearby galaxies at H$\alpha$, ultraviolet, FIR, and radio wavelengths.   
The H$\alpha$-SFR conversion they use, from \cite{Rosa2002},  is 40\% higher 
than \cite{Kennicutt98} in the sense that the derived Grimm SFRs would be higher using the \cite{Rosa2002} value.  
The radio conversion they use is from
\cite{Condon1992}, but they use the thermal radio-SFR conversion factor rather than the non-thermal (which are
different by an order of magnitude).   There appear to be relatively few radio measurements in the Grimm paper,
so this appears not to matter.   Thus it appears that H$\alpha$ measurements could result in SFR measurements
that are up to 40\% higher than the other authors, but these are not always used in their estimates.

\subsection{\boldmath$Chandra$-SDSS source matching}

In the 53 ACIS fields there are 338 ``well-observed" SDSS DR2 galaxies with
$0.03 \leq z \leq 0.25$.
Our definition of well-observed is that $Chandra$ covered at least 95\% of the galaxy's 50\%-encircled
light radius (we use the Petrosian 50\% encircled-light radius in the $g$-band) and that the galaxy was observed
at a $Chandra$ off-axis angle $<12^{\prime}$.  Including galaxies with off-axis angles larger than $12^{\prime}$ resulted in 
few additional detections.

\begin{figure}[t!]
\centerline{\includegraphics[scale=0.50,angle=0]{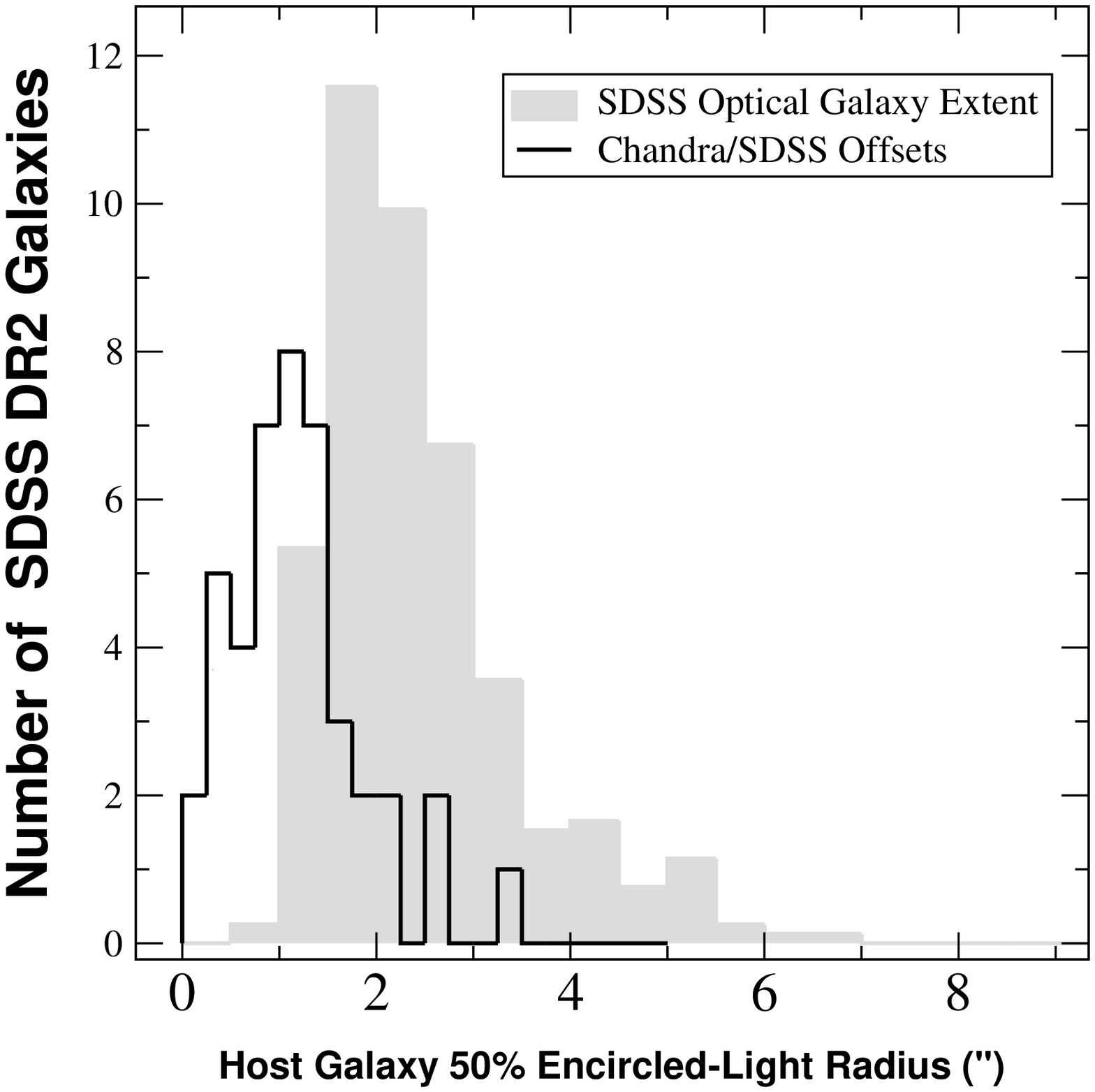}
\includegraphics[scale=0.50,angle=0]{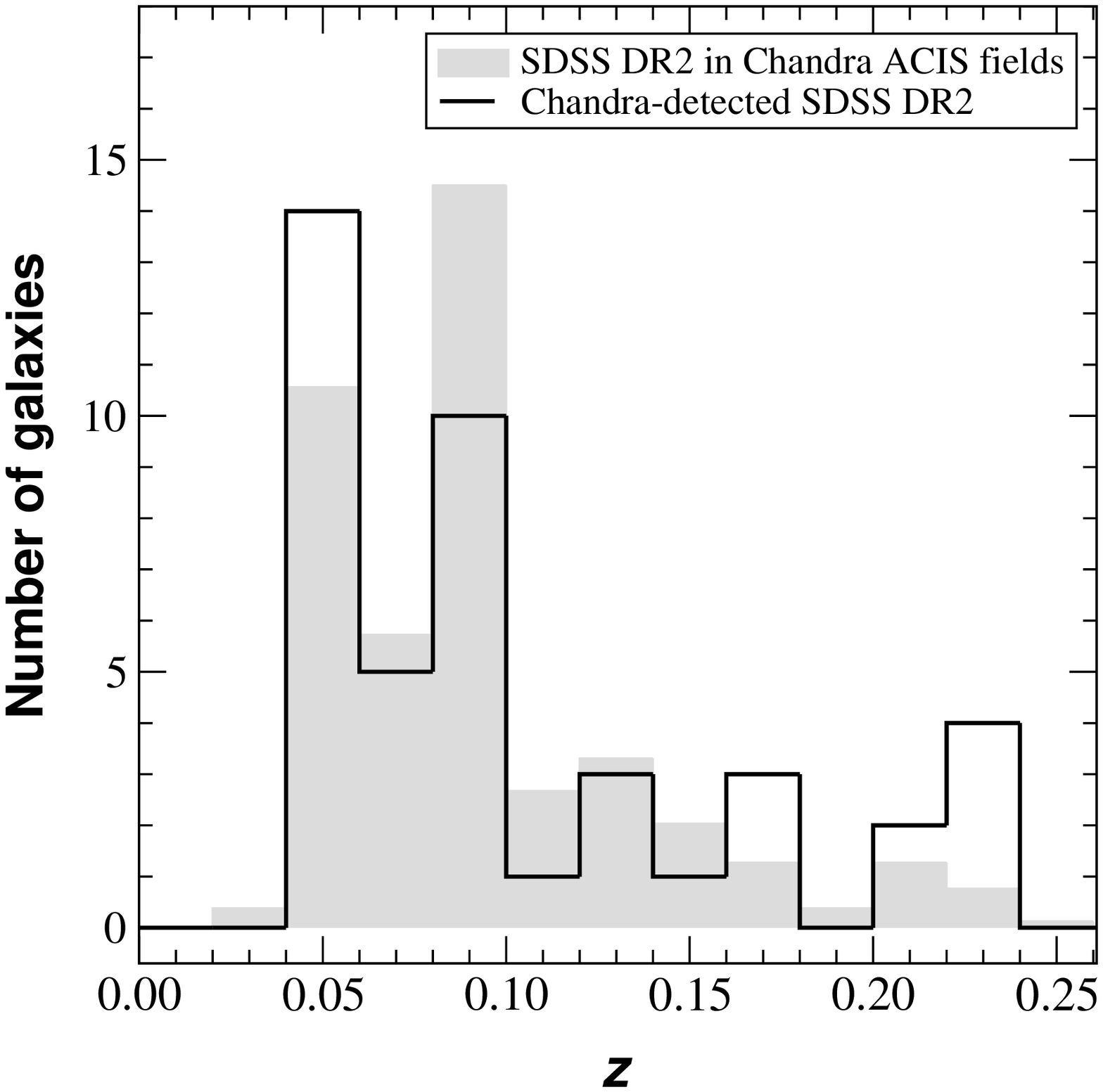} }
\caption[]{\label{zdistribution} {\bf (Left)} Comparison of the optical extent of
the galaxies with the X-ray/SDSS offsets.  Both are in units of arcseconds.
{\bf (Right)} Redshift distribution of the 42 serendipitously-detected SDSS DR2 galaxies in $Chandra$
ACIS fields (black line) as compared to the full X-ray observed sample of 338 SDSS DR2 galaxies (filled gray histogram).  The
full X-ray observed sample has been renormalized for ease of comparison.    
The properties of these 42 galaxies are given in Table~\ref{datatables}.  
}
\end{figure}

\begin{figure}[t!]
\centerline{\includegraphics[scale=0.80,angle=0]{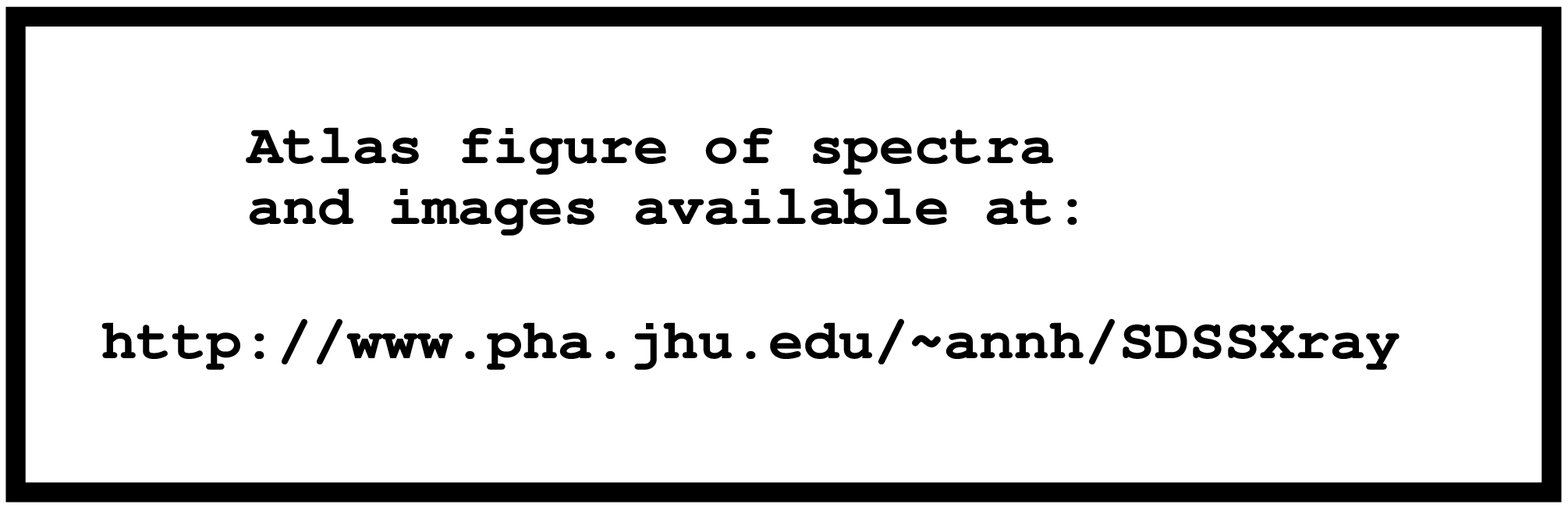}}
\caption[]{\label{images} Optical images and spectra for all the SDSS-$Chandra$ matches.
The red circle is the approximate position of the fiber aperture and the SDSS Right Ascension
and Declination is given in the upper left corner of each image.
}
\end{figure}

The SDSS astrometry is excellent ($<0$\farcs5) and as mentioned in \S~\ref{xrdata}, the
X-ray positions from {\sc XASSIST} are accurate to better than $2\farcs0$.   The optical extent of the SDSS galaxies was typically
larger than either of these two positional errors (the median galaxy radius was $\approx2$\farcs2; see Figure~\ref{zdistribution}a).
SDSS galaxies were considered matched to an X-ray source if the X-ray source position was within the optical 50\% encircled-light
radius (we used the Petrosian radius in the $g$-band).  We found that 42 galaxies contained X-ray detections and the X-ray detections
were typically well within the optical galaxy extent (see Figure~\ref{zdistribution}a).  
None of these 42 sources were the primary target of the X-ray observation but there is
clearly some bias towards cluster members, as indicated in the
redshift distribution in Figure~\ref{zdistribution}b.   
We randomly offset the SDSS galaxy source positions by 20$^{\prime \prime}$ and 30$^{\prime \prime}$ to evaluate the 
false-match probability and found that $<1$ false match is expected among the 42 galaxies. 
The X-ray and optical properties of these 42 galaxies are given in Table~\ref{datatables}, including the X-ray fluxes,
redshifts, absolute magnitudes, and optical spectroscopic classifications.  The X-ray source SDSS spectra and images are shown in Figure~\ref{images}.
The median number of 0.5--8~keV X-ray counts per source is just 18.2, so we do not present X-ray spectroscopic results in this paper.

We have examined
these sources carefully to ascertain that they are not affected by instrumental problems such
as proximity to a chip gap in the ACIS camera and/or proximity to any bright X-ray source.
The Galactic column density along the various
lines of sight to these galaxies are fairly low, the maximum \nh ~is $=4.7\times10^{20}$~\cmsq , the 
mean and standard deviation is \nh$=2.6\pm1.2\times10^{20}$~\cmsq.  X-ray fluxes are corrected for
Galactic absorption. 

There are published results on 37 X-ray-detected normal and starburst galaxies
in the CDF-N \citep{HornOBXF} so we again used this field to test our results.  These 37 CDF-N X-ray detected
galaxies are typically at higher redshifts and thus below the optical flux limit of SDSS (median $R\approx19$~mag). 
Three of the four SDSS DR2 galaxies in this field were X-ray undetected in our sample
(which did not use the full combined depth of the CDF-N).  One additional source is detected
in the full 2~Ms depth of \citep{HornOBXF} and the final
two were at large off-axis angles where the effective exposure time is low.  This illustrates that the extreme
depth of the CDF-N is not necessary for this kind of lower redshift work and also shows the
necessity of building up large areal coverage to probe the relatively sparse galaxy population at $z\approx0.1$.

\section{Results}

\subsection{Optical Galaxy Classification}

The majority of the $Chandra$-detected SDSS DR2 galaxies are ``emission-line" galaxies but the precise number
depends upon the significance cutoff used for the emission line flux.   Note that 
all classifications are based upon continuum-subtracted spectra and that the optical spectroscopic apertures typically
cover $\approx30$\% of the host galaxy.  

\begin{figure}[t!]
\centerline{\includegraphics[scale=0.65,angle=0]{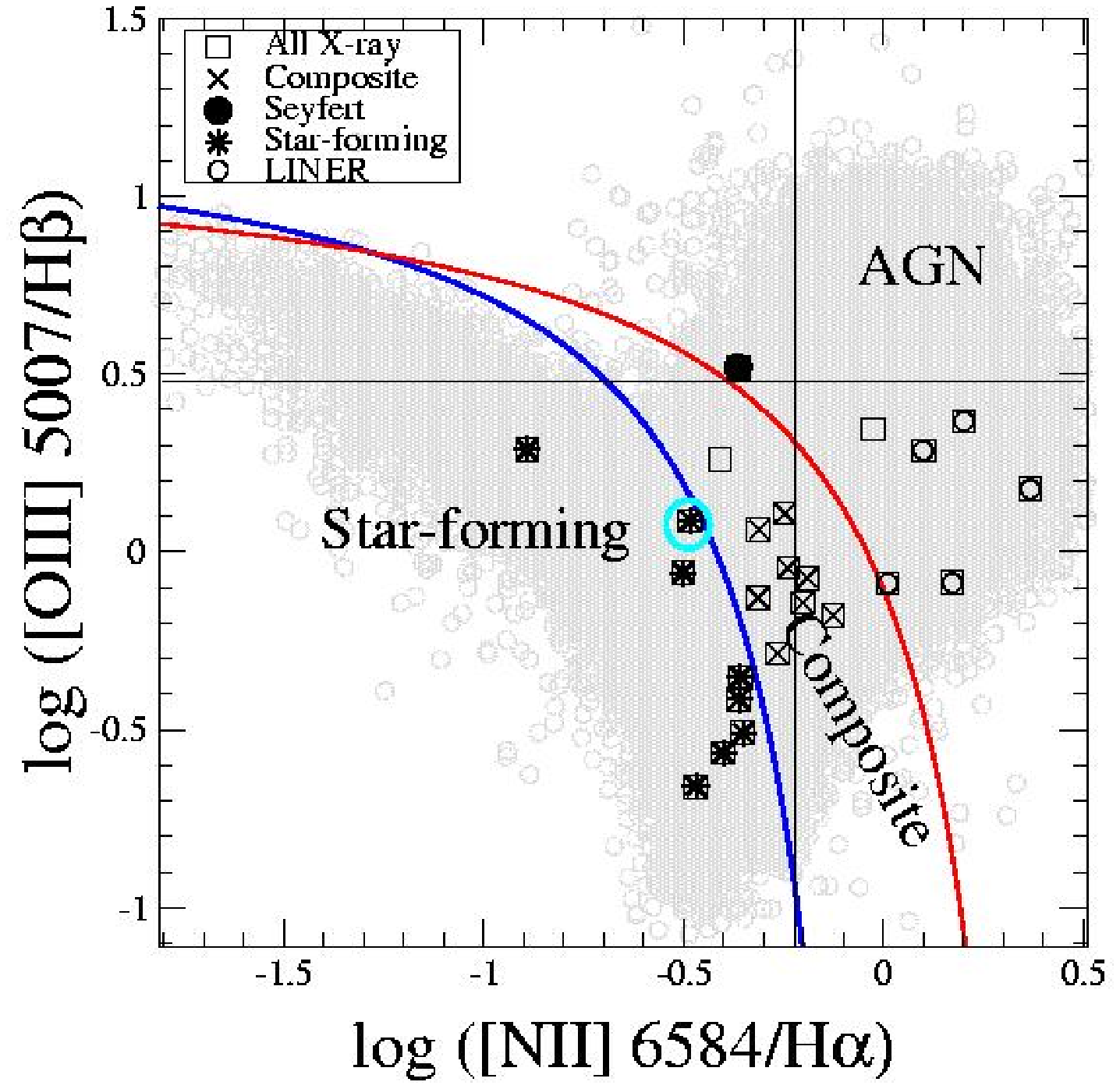}
\includegraphics[scale=0.45,angle=0]{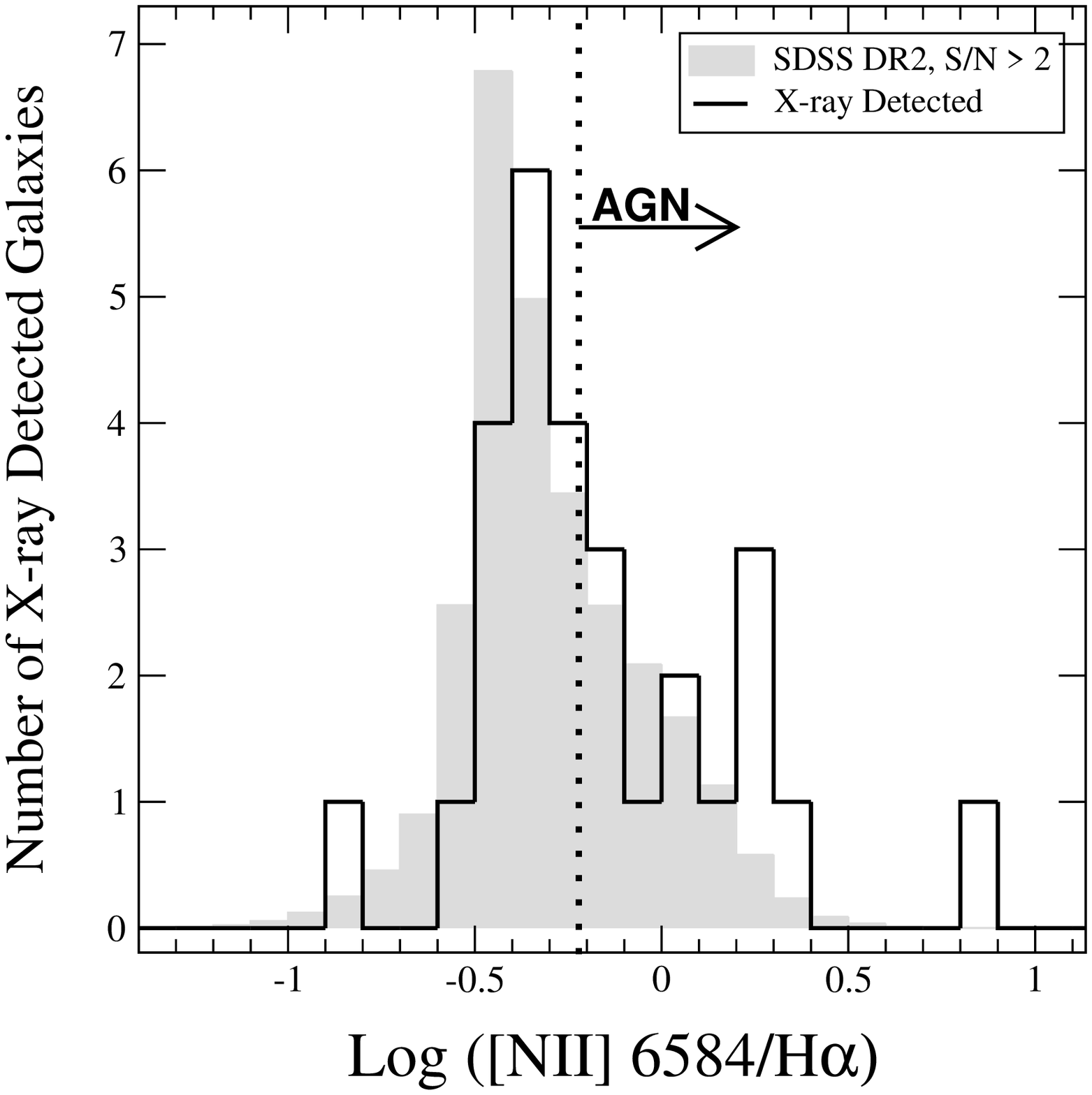}}
\caption[]{\label{emissionlineratios} {\bf (Left)} Emission-line ratios for the total SDSS DR2
sample (Gray circles) and the X-ray detected sources (black symbols).  Empty symbols indicate
sources with lower-significance emission lines ($>2\sigma$) whose classifications are thus 
slightly more uncertain.  One slightly lower significance source is off this diagram at
$ \biggl[ \log{([NII]/H\alpha)},\log{([OIII]/H\beta)} \biggr] = [ 0.8, 0.4 ] $.   The breakdown
of source classifications is given in Table~\ref{ERtypes}.  The source marked with
a cyan symbol is a Narrow-Line Seyfert 1 galaxy.   The vertical line indicates the
${[NII]/H\alpha} = 0.6$ demarcation for AGN vs. starbursts and the horizontal line
indicates the ${[OIII]/H\beta}=3$ demarcation for Seyferts vs. LINERs.
{\bf (Right)} $\log{([NII]/H\alpha)}$ for the 28 X-ray detected galaxies with both $[NII]$
and H$\alpha$ detected above 2$\sigma$ (this is a gain of an additional 3 emission-line galaxies).
There are 13 galaxies with ${[NII]/H\alpha} > 0.6$, versus 10 when using both ${[OIII]/H\beta}$
and ${[NII]/H\alpha}$. 

}

\end{figure}

The emission-line ratios for both the X-ray detected galaxies and the full SDSS DR2 galaxy 
sample are shown in Figure~\ref{emissionlineratios}a and their classifications based upon those ratios
are given in Table~\ref{ERtypes}.  Galaxies are classified if they have 
all four emission lines ([OIII], H$\beta$, [NII], H$\alpha$) detected at $>2\sigma$; there are 24 galaxies matching this criteria (see Table~\ref{ERtypes}).
It is also possible to use the $[NII]/H\alpha$ ratio alone (see Figure~\ref{emissionlineratios}b) as a coarse AGN discriminator; this yields an
additional 3 emission-line galaxies, {\it all} of which are classified as AGN.  Note that at high-redshift,
H$\alpha$ would be unobservable so these galaxies would likely have been misclassified.

There are thus 24 emission-line galaxies (ELGs), three lower-significance ELGs (with only 
[NII] and H$\alpha$ detected above $2\sigma$),
and 15 galaxies which appear to be absorption-dominated to fairly strict limits.

\subsection{X-ray/SFR Correlations}
\label{XRSFRsection}

\begin{figure}[t!]
\centerline{\includegraphics[scale=0.90,angle=0]{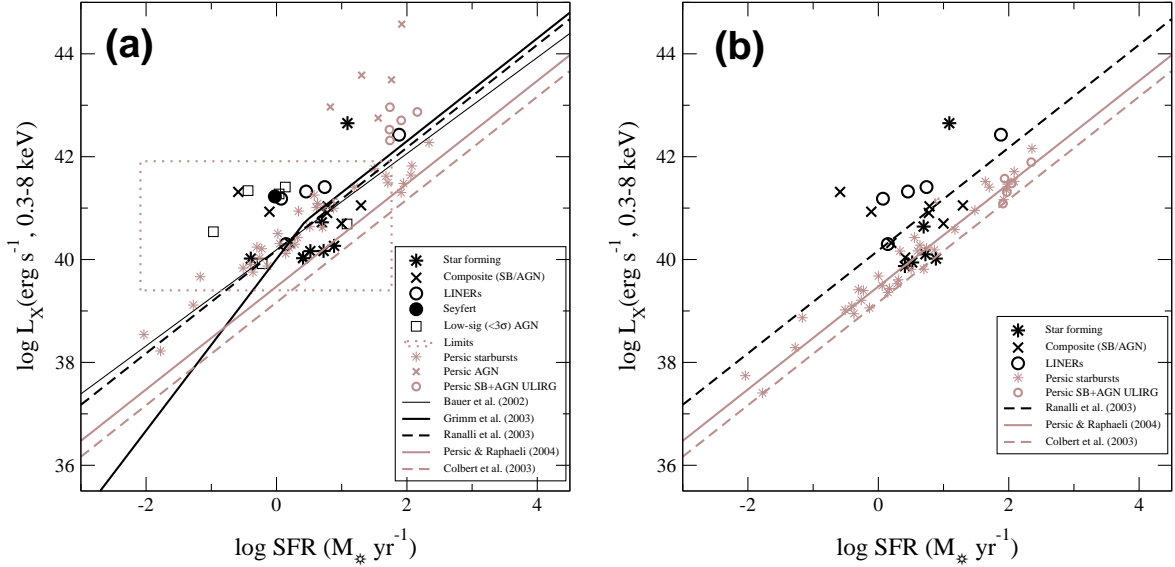}}
\caption[]{\label{SFRLX} {\bf (a)} Total $Chandra$ full-band X-ray luminosity versus star-formation rate 
(SFR) as calculated in \cite{Brinch2004}.  The lines show the correlations derived in several studies
in the literature, all of these studies have been adjusted to the 0.3--8~keV band assuming 
a $\Gamma=1.8$ power law spectrum and SFRs have been corrected to a Kroupa IMF.  We have divided the Colbert line by
a factor of 1.4 to remove their bolometric correction (see \S~\ref{IMFcomp}).
We also show the data on nearby galaxies from \cite{Persic2004}. 
The dashed box shows the region of the upper limits for the 295 X-ray-undetected SDSS galaxies, 
which are fairly evenly distributed throughout this area (plotting them individually confuses the plot).  
{\bf (b) }  Same as (a) but now the X-ray emission that is not associated with current star-formation has been removed
from the full-band X-ray luminosity by accounting for the LMXB component through stellar mass estimates.   Note that one ``star-forming"
galaxy is not shown in panel (b) because its entire X-ray luminosity is explained by the non-SFR component (appear near logLX,logSFR $=[40,-0.3]$ in panel (a)) For the SDSS DR2 
datapoints we have used the mass/LMXB coefficient from \cite{ColbertXLF2003} and subtracted this value from the total ${\rm L}_{\rm X}$.
 The \cite{Persic2004} data points in this graph are now the ``corrected" data points in Figure~4 of their paper, which has had a 
correction factor applied for the LMXB component; note that the difference is greatest for the lowest SFR.  
}

\end{figure}

In Figure~\ref{SFRLX} we show the total integrated X-ray luminosity in the 0.3--8~keV band 
for the $Chandra$-detected ELGs plotted against SFR as measured by 
\cite{Brinch2004}.   We also show the correlations between SFR and X-ray 
emission found by several other authors.  We have adjusted these various
correlations to the 0.3--8~keV bandpass. We assumed a $\Gamma=1.8$ power-law for conversion;
the correction is a factor of $\approx2$ from the 2--8~keV bandpass
to the 0.3--8~keV bandpass.   In Section~\ref{IMFcomp} we describe the various SFR measurements
used by these other authors; their methods yield differences of up to 50\% in SFR.  In Figure~\ref{SFRLX}
we divide the SFRs from the comparison studies (Colbert, Persic, Grimm, etc.) by a factor of 1.5 to account for the Kroupa IMF.
The Colbert SFRs were additionally divided by a factor of 1.4 to remove the bolometric correction over SFRs derived
purely from IRAS fluxes.

\begin{figure}[t!]
\centerline{\includegraphics[scale=0.90,angle=0]{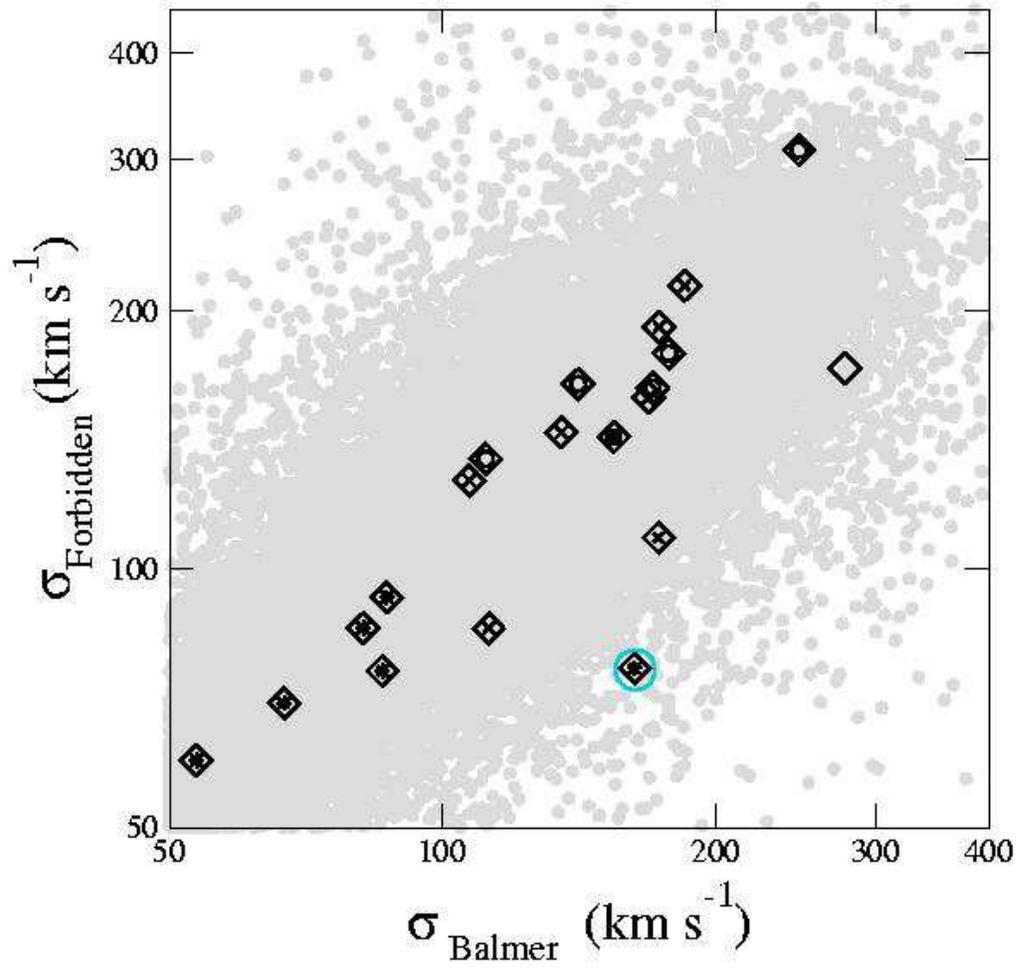}}
\caption[]{\label{BalmerForbidden} $\sigma_{\rm Balmer}$ versus $\sigma_{\rm Forbidden}$ for the full SDSS DR2 sample having
emission lines detected above 2$\sigma$ and for the X-ray detected emission-line galaxies (classification symbols the same
as in previous figures).  The NLS1, X102348.48+040552.4, is marked in cyan as in Figure~\ref{emissionlineratios}.  }
\end{figure}

There are seven pure ``star-forming" galaxies in our sample based upon emission-line ratio diagnostics alone.
There is one clear outlier among the X-ray-detected ``star-forming" galaxies in Figure~\ref{SFRLX} (X102348.48+040552.4; it is marked
in cyan in the BPT diagram of Figure~\ref{emissionlineratios}) 
that has a much higher X-ray luminosity than is expected based upon its SFR.  Inspection of its optical spectrum
reveals that it is a Narrow-Line Seyfert 1 (NLS1; see Figure~\ref{images}).   If one compares the width of the Balmer lines in this spectrum
with the widths of the forbidden lines, the Balmer width is found to be much larger.   Thus while optical emission
line ratio diagnostics are fairly good at separating AGN from star-forming galaxies, we find that the additional
information afforded by line widths is quite critical.  We have plotted the Balmer vs. forbidden line widths
for all the emission-line galaxies in Figure~\ref{BalmerForbidden}.   As can be seen, there are only three other sources
with Balmer widths that are appreciably larger than the forbidden line widths; these three are composite galaxies (or deemed to
be low signal-to-noise AGN).  

Correcting for the likely LMXB contribution using the stellar mass estimates from \cite{Kauff2003AGN} and the LMXB mass constant of \cite{ColbertXLF2003}  brings five of the six remaining
star-forming galaxies into agreement with the \cite{Persic2004} line;  one of these has such a low SFR that all of its 
X-ray emission is explained by LMXB emission (it thus does not appear in the second panel of Figure~\ref{SFRLX}.   
There remains only one ``star-forming" galaxy besides the NLS1 that has a higher value of X-ray emission than expected from the Colbert relation.  
This galaxy, X150944.45+570434.5, is in a particularly clean $Chandra$ field (there is a $z\approx4$ QSO in the field, but no
large scale structure, etc.).     The source has no detected 2-8~keV counts and thus does not show evidence for harboring an AGN.  

\subsection{Galaxies with Absorption-line Dominated Optical Spectra}

\begin{figure}[t!]
\centerline{\includegraphics[scale=0.70,angle=0]{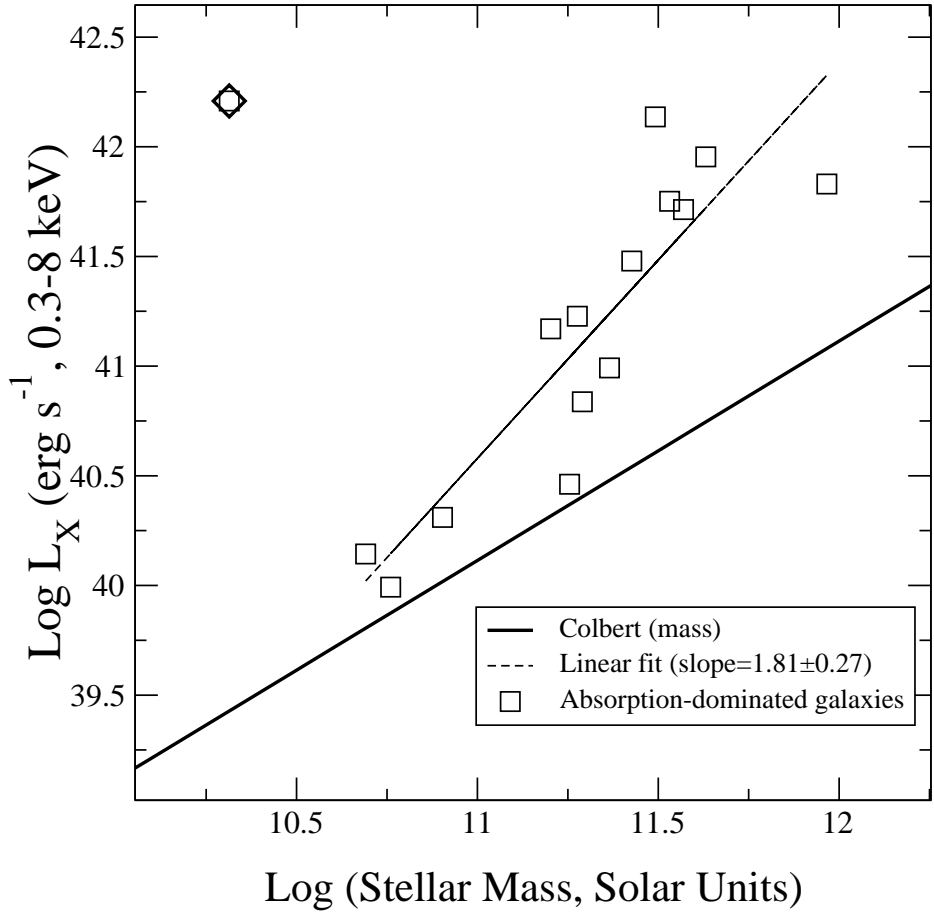}}
\caption[]{\label{MASSLX} Full-band X-ray luminosity versus stellar 
mass calculated in \cite{Kauff2003mass} for the 15 ``pure" absorption-dominated
galaxies.  These are the galaxies which do not demonstrate emission lines in
any of the four ``BPT diagram" lines (see Figure~\ref{emissionlineratios}) above 2$\sigma$
in the continuum-subtracted spectra.  These are very strong limits as these diagnostic
spectra cover the inner 30\% of the host galaxies and thus suffer much less dilution than
do higher-redshift samples.    The dark solid line indicates the stellar-mass/X-ray {\it point source}
relation of \cite{ColbertXLF2003} which is the expected LMXB component in these galaxies. 
The dashed line indicates a fit to the data, excluding the outlier X171206.83+640830.7 (marked with a diamond).
The slope of this line is consistent with the non-linear $L_{\rm X}$-$L_{\rm B}$ relation found by
\cite{Fabbiano92}, with slope $\approx1.8$.  The non-linearity of the relation was attributed to
the hot ISM of elliptical galaxies.}
\end{figure}

There are 15 galaxies which lack emission lines at the 2$\sigma$ level.   Expecting that these galaxies do not
have much current star-formation, we plot their {\it total} X-ray luminosities versus their host galaxies' stellar masses 
\citep[Figure~\ref{MASSLX}; the masses are the medians of the probability 
distributions of dust-corrected mass for each galaxy from][]{Kauff2003mass}.   The total X-ray luminosities exceed that expected from
X-ray binaries alone.  The $L_{\rm X}$-mass
relation is consistent with the $L_{\rm X}$--$L_{\rm B}$ relation of \cite{Fabbiano92} which was attributed to the
X-ray emission from hot gas in galaxies.

\begin{figure}[t!]
\centerline{\includegraphics[scale=0.50,angle=270]{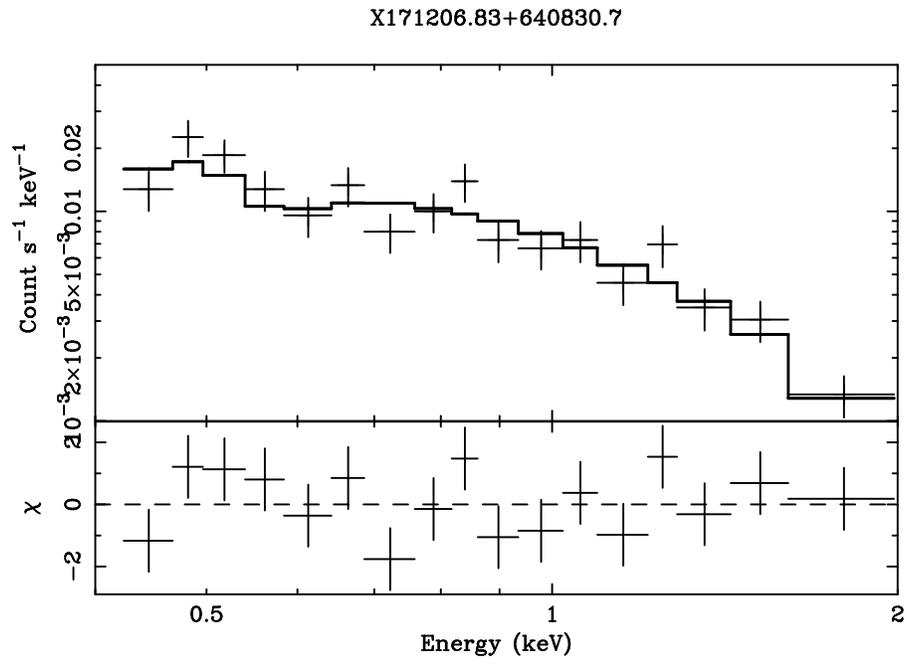}}
\caption[]{\label{XBONGxr} X-ray spectrum of the outlier X171206.83+640830.7 from Figure~\ref{MASSLX}. }
\end{figure}

The spectrum of the one outlier in the absorption-dominated galaxy X-ray/mass plot, X171206.83 $+$640830.7, does not show any
sign of AGN emission (see Figure~\ref{images}) and has been previously reported as a post-starburst galaxy by \cite{Davis2003}.   
The X-ray emission is extremely soft ($\Gamma\approx3.4$, see Figure~\ref{XBONGxr}) and the X-ray luminosity is quite high ($L_{\rm X}\approx1.6\times10^{42}$~\lumin )
but the source is detected in the 2--8~keV band as well with  $L_{\rm X} \approx 1.3\times10^{41}$~\lumin (2--8~keV).
There is no detected SFR within the aperture (no H$\alpha$ emission) and the estimated {\it current} SFR for the entire galaxy 
is 0.6~\msun~yr$^{-1}$, with a 97.5\%
upper limit at $4.1$~\msun~yr$^{-1}$ \citep{Brinch2004}.   The contribution from either an evolved component or {\it current}
star-formation to the hard X-ray luminosity is estimated to be $<5\%$.  
 \cite{Davis2003} indicate that this source may be variable, having declined in soft X-ray flux by a factor of several since $ROSAT$
observations in 1993 and 1994; they therefore suggest that this source is either an AGN or a bright Ultra-Luminous
X-ray source \citep[ULX; e.g., ][]{Colbert2002}.   Our continuum-subtracted
optical spectra place strong constraints on the AGN nature of this object (we do not detect any AGN signature) and if the lack of emission lines were due to obscuration, we might have expected to see evidence in the X-ray spectrum, which we do not (see Figure~\ref{XBONGxr}). So we explore the
post-starburst nature of this object in more detail.

\cite{Kauff2003mass} calculate burst mass fractions for such galaxies.
The median of the mass fraction distribution indicates that 10\% of the mass was produced in a burst, corresponding to 
$\approx 2\times10^{9}$\msun.  However, as noted in \cite{Kauff2003mass},
there is an age/mass degeneracy: a large burst that occurred long ago is indistinguishable from a smaller burst that occurred more recently. 
The distribution of the mass fraction actually extends from zero to 65\%, the zero case corresponds to the stellar mass having been 
formed a long time ago ($>2$Gyr).   We adopt the median value of 10\% acknowledging that there is some significant uncertainty.

\cite{Sipior2004} model the X-ray binary population resulting from a 10~\msun~yr$^{-1}$ starburst 
lasting 20 Myr; the 2--10 keV luminosity
of such a burst has a peak near $2 \times 10^{40}$~\lumin ~10~Myr after the burst has ended (assuming a Salpeter IMF), 
declining to $<10^{38}$~\lumin by 1~Gyr after
the burst.  If X171206.83+640830.7 were a post-starburst galaxy with 10\% of its mass formed in a 20~Myr interval, 
the implied star-formation would be 150~\msun~yr$^{-1}$ (scaling by a factor of 1.5 to account for the Kroupa IMF).  
Assuming we observe X171206.83+640830.7 very soon after the burst (near the X-ray maximum at burst+10~Myr) then scaling the
\cite{Sipior2004} curves yields an SFR of $\approx 65$~\msun~yr$^{-1}$, which within errors is 
very consistent.    We
find the coarse agreement between likely post-starburst nature of
this galaxy and the X-ray emission encouraging.

However, the lack of emission lines indicate that this source must be viewed much more than 10~Myr after
its peak in X-ray luminosity.  Since the peak X-ray luminosity would be even higher than observed now, 
the implied evolved starburst strength would be scaled upward by nearly an order of magnitude.  
A more plausible explanation for the high X-ray luminosity, given the X-ray variability, 
is that this is an extremely X-ray luminous accreting binary.  Such a source would thus be 
the most X-ray luminous Ultraluminous X-ray source \citep[ULX; see e.g., ][]{Fabbiano03} ever detected.

\subsection{AGN Activity and XBONGS}

There are 19 galaxies in this sample which may be classified as AGN (see Table~\ref{ERtypes}).  Their
  $L_{\rm X}/L_{[OIII]}$ ratios span the range found for nearby Seyfert 2 galaxies \citep[e.g.,][]{Papa2001}. About 12 of
the galaxies are at the upper end of the expected range, indicating possible Compton-thick obscuration 
($L_{\rm X}/L_{[OIII]} > 10$).   The statistics on the hardness ratios for these 12 objects
[ratio of (H-S)/(H+S) where H=2--8~keV counts and S=0.5--2~keV soft counts] are not sufficient to test for a correlations with $L_{\rm X}/L_{[OIII]}$ 
\citep[as was done for a nearby sample of Seyferts in ][]{Papa2001}.   We will be exploring these high $L_{\rm X}/L_{[OIII]}$ AGN
in more detail in a future paper (A. Hornschemeier et~al., in preparation). 


There are three AGN with no detected [OIII]/H$\beta$ but H$\alpha$/[NII] ratios indicating they are AGN.
These three sources are candidate ``XBONGS" (X-ray Bright, Optically Normal Galaxies) which is an observational class of AGN 
found in deep X-ray surveys with no optical signature for AGN activity.  Similar objects at higher redshift (in deeper surveys) would likely
have been classified as absorption-dominated galaxies due to the inaccessibility of
the H$\alpha$ and [NII] emission lines.
These three candidate XBONG galaxies do not, however, have high X-ray luminosities ($L_{\rm X} < 2 \times 10^{41}$~\lumin, 0.3--8~keV). 
Thus they do not appear to be the observational analogs to XBONGS found in hard X-ray surveys 
\citep[e.g.,][]{Comastri02}.  There are similarly no extremely hard, X-ray luminous sources among the 15 X-ray detected 
absorption-dominated X-ray-detected galaxies.  There are a total of 103 absorption-dominated galaxies in all of the $Chandra$ fields at
off-axis angles $<8^{\prime}$ (13 of the 15 X-ray detected absorption-line galaxies are at off-axis angle $<8^{\prime}$.  Thus we do not 
find any galaxies which show signs for extremely luminous AGN in the X-ray band that show no optical signature of activity in the high 
signal-to-noise SDSS spectra.

\section{Conclusions}

We have conducted a pilot study of 42 SDSS DR2 galaxies serendipitously detected in the X-ray band.  The high
signal-to-noise SDSS spectra afford detailed activity diagnostics that are often missing in the deepest surveys.

Among X-ray detected galaxies which show no signs for AGN activity (15 absorption-dominated galaxies 
and 7 star-forming galaxies), the X-ray emission is naturally explained by a combination of LMXBs, 
HMXBs, and hot ISM.  Thus, we do not find evidence for AGN that are missed in the SDSS spectroscopic 
analysis.  Our work thus supports that of others \citep[e.g.,][]{Moran2002}
that fairly simple observational effects may make X-ray detected AGN appear as XBONGS at higher-redshift.
Among the 19 SDSS galaxies which have AGN signatures in their optical spectra, the $L_{\rm X}$-$L_{[OIII]}$
ratios are found to be coarsely consistent with Seyfert 2's in the nearby Universe.  

We do find four instances where galaxies might have been misclassified in lower signal-to-noise optical spectra.  A NLS1's line widths
clearly reveal the presence of an AGN while its emission-line ratios are similar to that of a starburst galaxy.   Three other
galaxies have pure absorption-dominated optical spectra blueward of H$\alpha$ but exhibit H$\alpha$ and [NII] emission typical of
AGN.  All three of these galaxies would not have been identified as AGN above $z\approx0.5$.  

The detection of X-ray emission from a post-starburst galaxy indicated that the burst mass fraction calculated from the optical
spectroscopic models may be in agreement with the expected X-ray emission from a starburst.  However, the likely
evolved nature of the post-starburst (the age of which is poorly constrained) indicates that the X-ray luminosity
is quite excessive given the magnitude of the starburst. This one object is not sufficient to 
test the models but is strongly suggestive that a large sample of post-starburst galaxies (perhaps with stronger constraints on the
burst mass fractions) may have a strong impact on the modeling of accreting binary systems.  It may also be 
an excellent way to search for young ULX sources as this galaxy appears to harbor a variable X-ray source.  We are able to rule
out that this post-starburst galaxy is an AGN to very strong limits from optical spectroscopy.

We have made an effort to compare the various $L_{\rm X}$-SFR relationships in the literature. We find that differences among e.g.,
bolometric corrections and IMF account for up to a 50\% difference in the SFR relationships.    Any difference greater than this level
must then be attributed to e.g., a different treatment of the contributions of AGN vs. star-formation. 
Our analysis supports the lower normalization of the $L_{\rm X}$-SFR relation after proper account of AGN activity and contributions
from evolved stellar populations are made.   

\acknowledgments

We gratefully acknowledge the constructive comments of an anonymous referee
through which the manuscript was significantly improved.
We are grateful to Guinevere Kauffmann for providing
useful comments on draft versions of this paper.
AEH gratefully acknowledges the financial support of
$Chandra$ fellowship grant PF2-30021.
CAT acknowledges support from NASA grant NAG~58426 and NSF grant
AST-0307386.  
Funding for the creation and distribution of the SDSS Archive has been
provided by the Alfred P. Sloan Foundation, the Participating
Institutions, the National Aeronautics and Space Administration, the
National Science Foundation, the U.S. Department of Energy, the
Japanese Monbukagakusho, and the Max Planck Society. The SDSS Web site
is http://www.sdss.org/.
The SDSS is managed by the Astrophysical Research Consortium (ARC) for
the Participating Institutions. The Participating Institutions are The
University of Chicago, Fermilab, the Institute for Advanced Study, the
Japan Participation Group, The Johns Hopkins University, Los Alamos
National Laboratory, the Max-Planck-Institute for Astronomy (MPIA),
the Max-Planck-Institute for Astrophysics (MPA), New Mexico State
University, University of Pittsburgh, Princeton University, the United
States Naval Observatory, and the University of Washington.
 This research has made use of the NASA/IPAC Extragalactic Database which is 
operated by JPL under contract with NASA.
%


\bibliographystyle{apj}
\bibliography{Hornschemeier}


\begin{deluxetable}{rrcrrrrrrrcrr}
\rotate
\tablecolumns{10}
\tabletypesize{\footnotesize}
\tablewidth{0pt}
\tablecaption {SDSS DR2 Galaxies Detected Serendipitously by $Chandra$ \label{datatables} }
%
%
\tablehead{
\colhead{XASSIST ID}  & 
\colhead{ $\Delta$SDSS-XR} &
\colhead{ Counts/Error} &
\colhead{ 0.3--8~keV Flux} &
\colhead{ Redshift}  & 
\colhead{M$_{z}^{\rm a}$   }  & 
\colhead{ $\log $SFR$^{\rm b}$}  & 
\colhead{ [ OIII ] Lum.} &   
\colhead{ $\log($mass$)^{\rm c}$  }  & 
\colhead{Type} \\
\colhead{}  &
\colhead{($^{\prime \prime}$) }  &
\colhead{ } &
\colhead{ (erg cm$^{-2}$ s$^{-1}$) } &
\colhead{ }  &
\colhead{ }  &
\colhead{ (\msun yr$^{-1}$) }  &
\colhead{ (\lsun) }  &
\colhead{ (\msun) }  &
\colhead{}  \\
}
\startdata
   X004146.77-092313.0 &   0.2 &  18.4$\pm  5.0$ &    4.2 &  0.059 &  -21.72 &   -0.97 & \nodata &    10.9 &       AGN(H$\alpha$ only)\\
   X011238.24+152804.5 &   1.3 &  10.2$\pm  3.7$ &    2.0 &  0.043 &  -21.36 &   -0.22 & \nodata &    10.6 &           AGN($<2\sigma$)\\
   X011253.91+153607.9 &   1.1 &  18.8$\pm  6.2$ &    3.5 &  0.041 &  -21.98 &   -0.94 & \nodata &    10.7 &                Absorption\\
   X011254.04+153248.9 &   0.3 &  22.1$\pm  6.1$ &    4.1 &  0.046 &  -22.57 &    0.17 & \nodata &    10.9 &                Absorption\\
   X011259.50+153528.4 &   1.4 &  18.9$\pm  4.6$ &    8.5 &  0.041 &  -21.78 &    0.41 &    6.21 &    10.9 &                 Composite\\
   X011313.86+152938.2 &   1.0 &  18.3$\pm  4.6$ &    3.3 &  0.230 &  -23.91 &    0.17 & \nodata &    11.6 &                Absorption\\
   X011319.63+152926.2 &   2.5 &  10.2$\pm  3.5$ &    1.9 &  0.049 &  -21.15 &    0.41 &    6.56 &    10.4 &              Star-forming\\
   X011454.23+003026.0 &   0.6 &  10.2$\pm  3.3$ &    2.2 &  0.044 &  -21.71 &   -0.94 & \nodata &    10.8 &                Absorption\\
   X011454.22+001812.1 &   0.7 &   7.7$\pm  3.0$ &    2.3 &  0.044 &  -22.31 &   -0.39 &    5.32 &    11.1 &              Star-forming\\
   X011508.79+001557.7 &   0.8 &  12.4$\pm  3.7$ &    3.4 &  0.043 &  -21.39 &    0.52 &    6.67 &    10.6 &              Star-forming\\
   X011515.78+001248.8 &   0.3 & 188.5$\pm 15.7$ &   53.7 &  0.045 &  -22.45 &    0.74 &    5.69 &    11.1 &                     LINER\\
   X011519.39+001357.2 &   0.3 &  15.4$\pm  4.1$ &    4.3 &  0.089 &  -21.59 &   -0.11 &    5.64 &    10.6 &                 Composite\\
   X011520.87+001534.1 &   0.6 &  16.7$\pm  4.2$ &    4.5 &  0.043 &  -21.84 &    0.15 &    6.07 &    10.8 &                     LINER\\
   X085224.96+511211.6 &   1.7 & 116.9$\pm 16.7$ &   12.9 &  0.073 &  -21.53 &   -0.03 &    7.15 &    10.6 &                   Seyfert\\
   X085232.43+512414.3 &   1.2 &  50.0$\pm 12.7$ &    7.4 &  0.125 &  -23.49 &   -0.17 & \nodata &    11.4 &                Absorption\\
   X093414.33+611735.7 &   1.7 &  44.4$\pm 12.8$ &   10.9 &  0.208 &  -23.71 &   -0.01 & \nodata &    11.5 &                Absorption\\
   X095249.26+515305.5 &   1.0 &  81.7$\pm 31.4$ &   19.7 &  0.215 &  -24.30 &    1.88 &    6.39 &    11.8 &                     LINER\\
   X102328.09+040917.3 &   1.4 &   6.3$\pm  2.6$ &    1.3 &  0.069 &  -21.11 &    0.72 &    6.58 &    10.2 &              Star-forming\\
   X102348.48+040552.4 &   1.4 & 791.7$\pm 28.9$ &  180.9 &  0.099 &  -22.08 &    1.09 &    6.93 &    10.9 &   ``Star-forming", NLS1$^{\rm d}$\\
   X103258.09+573106.4 &   0.8 &  13.9$\pm  4.5$ &    4.3 &  0.045 &  -21.01 &    0.20 &    5.22 &    10.3 &                 Composite\\
   X103259.76+575321.9 &   0.3 &  18.1$\pm  4.4$ &    5.3 &  0.123 &  -23.10 &    0.45 &    5.89 &    11.2 &                     LINER\\
   X111256.55+554056.9 &   3.4 &  17.4$\pm  5.5$ &    2.6 &  0.055 &  -22.05 &    0.88 &    5.87 &    10.8 &              Star-forming\\
   X122139.70+491953.7 &   2.2 &  14.6$\pm  4.1$ &    2.0 &  0.125 &  -22.60 &    0.78 &    6.53 &    10.9 &                 Composite\\
   X122930.53+032942.4 &   1.5 &  31.7$\pm  8.6$ &   10.1 &  0.078 &  -22.09 &    0.07 &    5.75 &    11.0 &                     LINER\\
   X123649.03+620439.9 &   1.0 &  21.3$\pm  8.2$ &    1.5 &  0.113 &  -23.25 &    1.08 & \nodata &    11.4 &       AGN(H$\alpha$ only)\\
   X124240.63+024020.4 &   0.4 & 114.0$\pm 13.5$ &   12.0 &  0.086 &  -22.82 &   -0.44 & \nodata &    11.1 &           AGN($<2\sigma$)\\
   X143821.88+034013.2 &   0.2 &  23.6$\pm 10.4$ &    4.5 &  0.225 &  -24.66 &    1.72 & \nodata &    12.0 &                Absorption\\
   X143826.34+033859.4 &   0.7 &  19.1$\pm  5.0$ &    3.4 &  0.235 &  -23.75 &    0.28 & \nodata &    11.5 &                Absorption\\
   X150554.29+013535.5 &   1.1 &  10.6$\pm  3.5$ &    1.9 &  0.170 &  -23.48 &    0.97 & \nodata &    11.2 &                Absorption\\
   X150944.45+570434.5 &   1.0 &  11.9$\pm  4.0$ &    0.9 &  0.150 &  -22.35 &    0.69 &    6.51 &    10.8 &              Star-forming\\
   X163608.18+410505.4 &   1.8 &  11.3$\pm  3.9$ &    1.4 &  0.170 &  -22.71 &    0.78 &    7.89 &    11.2 &                 Composite\\
   X163615.46+405714.7 &   0.8 &  19.7$\pm  7.0$ &    2.4 &  0.169 &  -23.34 &    0.04 & \nodata &    11.5 &       AGN(H$\alpha$ only)\\
   X163629.56+410221.6 &   1.4 & 288.2$\pm 17.4$ &   39.0 &  0.047 &  -21.33 &   -0.59 &    6.18 &    10.5 &                 Composite\\
   X163953.65+465226.4 &   1.2 &  24.6$\pm  7.9$ &    6.0 &  0.225 &  -23.89 &   -0.14 & \nodata &    11.6 &                Absorption\\
   X170209.18+641220.7 &   2.0 &  30.1$\pm  6.1$ &    6.4 &  0.084 &  -23.04 &    1.30 &    8.04 &    11.2 &                 Composite\\
   X171206.83+640830.7 &   1.1 & 402.1$\pm 21.0$ &   96.5 &  0.082 &  -21.01 &   -0.21 & \nodata &    10.3 &                Absorption\\
   X171232.43+640008.7 &   1.3 &  16.6$\pm  4.5$ &    3.8 &  0.085 &  -22.82 &   -0.51 & \nodata &    11.3 &                Absorption\\
   X171254.28+635930.2 &   2.6 &   7.2$\pm  3.2$ &    1.7 &  0.083 &  -23.13 &   -0.40 & \nodata &    11.3 &                Absorption\\
   X171303.84+640700.7 &   1.1 &  23.7$\pm  6.2$ &    6.1 &  0.081 &  -22.85 &   -0.28 & \nodata &    11.4 &                Absorption\\
   X171329.16+640247.6 &   1.1 &  71.3$\pm 10.4$ &   17.0 &  0.078 &  -23.44 &    0.13 & \nodata &    11.5 &           AGN($<2\sigma$)\\
   X221722.73+002107.0 &   1.7 &  17.1$\pm  4.7$ &    2.2 &  0.095 &  -22.12 &    1.00 &    6.39 &    11.0 &                 Composite\\
   X234817.93+010615.2 &   2.1 &  18.7$\pm  6.4$ &    7.8 &  0.093 &  -23.30 &   -0.39 & \nodata &    11.3 &                Absorption\\
\enddata
\tablenotetext{a}{Absolute magnitude in the $z$-band from \cite{Kauff2003AGN}}
\tablenotetext{b}{Star-formation rate from \cite{Brinch2004}}
\tablenotetext{c}{The median of the dust-corrected stellar mass distribution from \cite{Kauff2003mass}}
\tablenotetext{d}{This source has emission-line ratios typical of a star-forming galaxy but it is in fact a NLS1.  See \S\ref{XRSFRsection}. }

\end{deluxetable}

\clearpage

\begin{deluxetable}{rcrrrr}
\tablewidth{0pt}
\tablecaption{\label{ERtypes} Optical Spectroscopic Type of Serendipitous Chandra Detections}
%
%
\tablehead{
\multicolumn{1}{c}{Optical spectral type$^{\rm a}$ } 		&
\multicolumn{3}{c}{Number of X-ray detected galaxies} 	\\
\multicolumn{1}{c}{} &
\multicolumn{1}{c}{$>3\sigma$ class$^{\rm b}$}		&
\multicolumn{1}{c}{$>2\sigma$ class$^{\rm b}$}              &
\multicolumn{1}{c}{H$\alpha$ only$^{\rm c}$}			&
}
\startdata
Star-forming & 7  & 7 & \\
Composite  & 8 & 9 \\
LINER      & 5 & 6 \\
Seyfert    & 1 & 1 \\
\hline
{\bf Total Emission Line:}  &  21 & 24  & 27$^{\rm c}$ \\ 
\hline
{\bf Absorption} & 21 & 18  & 15 \\
\hline
TOTAL	   & 42 & 42  & 42 \\
\enddata
\tablenotetext{a}{As determined from emission-line ratios only (see 
Figure~\ref{emissionlineratios}.}
\tablenotetext{b}{Classifications made for objects who had all four emission-lines
measured above this significance.  Thus, the ``absorption-dominated" galaxies are those
galaxies for which emission lines may be present, but at least one is below the 
N$\sigma$ value.}
\tablenotetext{c}{Emission-line classification based on only the $\log{[NII]/H\alpha}$ ratio. 
13 sources are AGN and 15 sources are star-forming galaxies.}
\end{deluxetable}


\end{document}